\documentclass[letterpaper,twocolumn,accepted=2022-01-28]{quantumarticle}
\pdfoutput=1
\usepackage[utf8]{inputenc}
\usepackage[english]{babel}
\usepackage[T1]{fontenc}
\usepackage[numbers,sort&compress]{natbib}
\usepackage{amsmath,amsfonts,amssymb,amsthm,bbm,graphicx,enumerate,times}
\usepackage{mathtools}
\usepackage[usenames,dvipsnames]{color}
\usepackage{hyperref}
\usepackage{graphicx}
\usepackage{float}
\usepackage{etoolbox}
\usepackage{accents}
\usepackage{stmaryrd}

\usepackage{dsfont}

 \renewcommand{\i}{\,\ensuremath\mathrm{i}}

\DeclareMathOperator{\arcosh}{arcosh}

\newcommand{\ket}[1]{\left.\left|{#1}\right.\right\rangle}

\DeclareMathAlphabet{\mathpzcc}{OT1}{pzc}{m}{it}
\DeclareMathAlphabet{\mathpzc}{T1}{pzc}{m}{it}{\huge}

\newcommand{\m}{\operatorname{\gamma}}

\def\and{\quad\text{and}\quad}

\begin{document}

\title{Tensor network models of AdS/qCFT}

\author{Alexander Jahn,$^{1,2}$ Zolt{\'a}n Zimbor{\'a}s,$^{3,4,5}$ and Jens Eisert$^{1,6}$}
\affiliation{$^1$Dahlem Center for Complex Quantum Systems, Freie Universit{\"a}t Berlin, 14195 Berlin, Germany \\
$^2$Institute for Quantum Information and Matter, California Institute of Technology, Pasadena, CA 91125, USA \\
$^3$Institute for Particle and Nuclear Physics, Wigner Research Centre for Physics, 1121 Budapest, Hungary \\
$^4$BME-MTA Lend\"ulet Quantum Information Theory Research Group, 1111 Budapest, Hungary\\
$^5$Faculty of Informatics, Eötvös Loránd University, 1117 Budapest, Hungary \\
$^6$Department of Mathematics and Computer Science, Freie Universit{\"a}t Berlin, 14195 Berlin, Germany
}

\begin{abstract}
The study of critical quantum many-body systems through conformal field theory (CFT) is one of the pillars of modern quantum physics. Certain CFTs are also understood to be dual to higher-dimensional theories of gravity via the anti-de Sitter/conformal field theory (AdS/CFT) correspondence.
To reproduce various features of AdS/CFT, a large number of discrete models based on tensor networks have been proposed. Some recent models, most notably including toy models of holographic quantum error correction, are constructed on regular time-slice discretizations of AdS.
In this work, we show that the symmetries of these models are well suited for approximating CFT states, as their geometry enforces a discrete subgroup of conformal symmetries. Based on these symmetries, we introduce the notion of a quasiperiodic conformal field theory (qCFT), a critical theory less restrictive than a full CFT and with characteristic multi-scale quasiperiodicity.
We discuss holographic code states and their renormalization group flow as specific implementations of a qCFT with fractional central charges and argue that their behavior generalizes to a large class of existing and future models.
Beyond approximating CFT properties, we show that these can be best understood as belonging to a paradigm of discrete holography.
\end{abstract}

\maketitle

\section{Introduction}

Quantum field theories constitute a central cornerstone of modern theoretical physics, describing a large part of physical phenomena from condensed matter to high-energy settings. 
From the start of the field, symmetries have played a central role in identifying and describing physical theories, perhaps most prominently in the case of \emph{conformal field theories} (CFTs) \cite{BigYellowBook1997,Ginsparg:1988ui,Blumenhagen:2009zz} which possess an invariance under local scale transformations. These theories, describing the physics of second-order phase transitions as well as being an integral part of the \emph{anti-de Sitter/conformal field theory correspondence} (AdS/CFT) \cite{Maldacena98,Witten:1998qj}, have properties that are highly restricted by their symmetries. Unlike the well-studied $1{+}1$-dimensional case, analytical techniques of continuum CFTs in higher dimensions are scarce and few exact results exist. To circumvent these issues, two popular approaches exist: First, AdS/CFT allows for many results of strongly interacting CFTs to be mapped to weakly interacting gravitational theories, usually under the assumption of additional symmetries. Second, one may study the properties of a \emph{discretized} CFT model numerically.
For this latter approach, tensor network models have shown particular promise: The \emph{multi-scale entanglement renormalization ansatz} (MERA) \cite{PhysRevLett.101.110501} has been shown to accurately reproduce a large range of CFT properties in $1{+}1$ dimensions \cite{Pfeifer:2008jt,Milstead:2018MERAconformal} and is extensible to higher dimensions \cite{Evenbly:2008pza}.
The MERA has also been proposed as a discrete realization of AdS/CFT  \cite{PhysRevD.86.065007,PhysRevD.97.026012} albeit possessing a geometry that does not directly match with that of its higher-dimensional gravitational theory \cite{Beny:2011vh,Bao:2015uaa,Milsted:2018san}. 
In a parallel development, tensor network approaches have incorporated \emph{holographic quantum error-correcting codes}, reproducing some properties of critical theories, yet lacking a clear connection to CFTs (for a recent review, see Ref.\ \cite{JahnReview}).

In this work, we show that tensor networks on regular hyperbolic tilings provide a natural CFT ansatz by extending ideas from continuum AdS/CFT to the discrete setting. Crucially, their boundary states obey a discrete subset of the continuum CFT symmetries \emph{directly from the tensor network geometry}, reproducing many of the properties of continuum CFT ground states.
Rather than breaking continuum translation invariance at a fixed lattice scale, this approach tolerates soft lattice effects appearing at all scales in a fractal manner, following a renormalization group (RG) prescription that allows fine- and coarse-graining to arbitrary scales.
Building on previous work that identified the boundary symmetries of these tilings with those of a \emph{quasicrystal} that suggested a notion of discrete conformal geometry \cite{Boyle:2018uiv}, we show how this picture can be modified and extended to describe tensor network states whose symmetries are captured by a \emph{multi-scale quasicrystal ansatz} (MQA) with contributions from all tiling layers, each corresponding to a length scale in the boundary state.
Further characterizing the resulting symmetries, we argue that these states realize instances of what we call a \emph{quasiperiodic conformal field theory} (qCFT) with inherently discrete structure that exhibits discrete analogues of scale and translation invariance found in a more constrained continuum CFT. These boundary transformations follow from invariance transformations of the hyperbolic disk that preserve regular hyperbolic tilings, characterized by Fuchsian groups.
As this discretization produces specific relationships between bulk and boundary symmetries, such tensor network models should be understood as realizations of an \emph{AdS/qCFT correspondence} between discrete theories rather than merely as an approximation of continuum AdS/CFT. Using examples from previous literature in which tensor network models with these symmetries have already been constructed, we illustrate the range of realizable qCFTs models.

\begin{figure*}[tb]
\centering

\includegraphics[height=0.188\textheight]{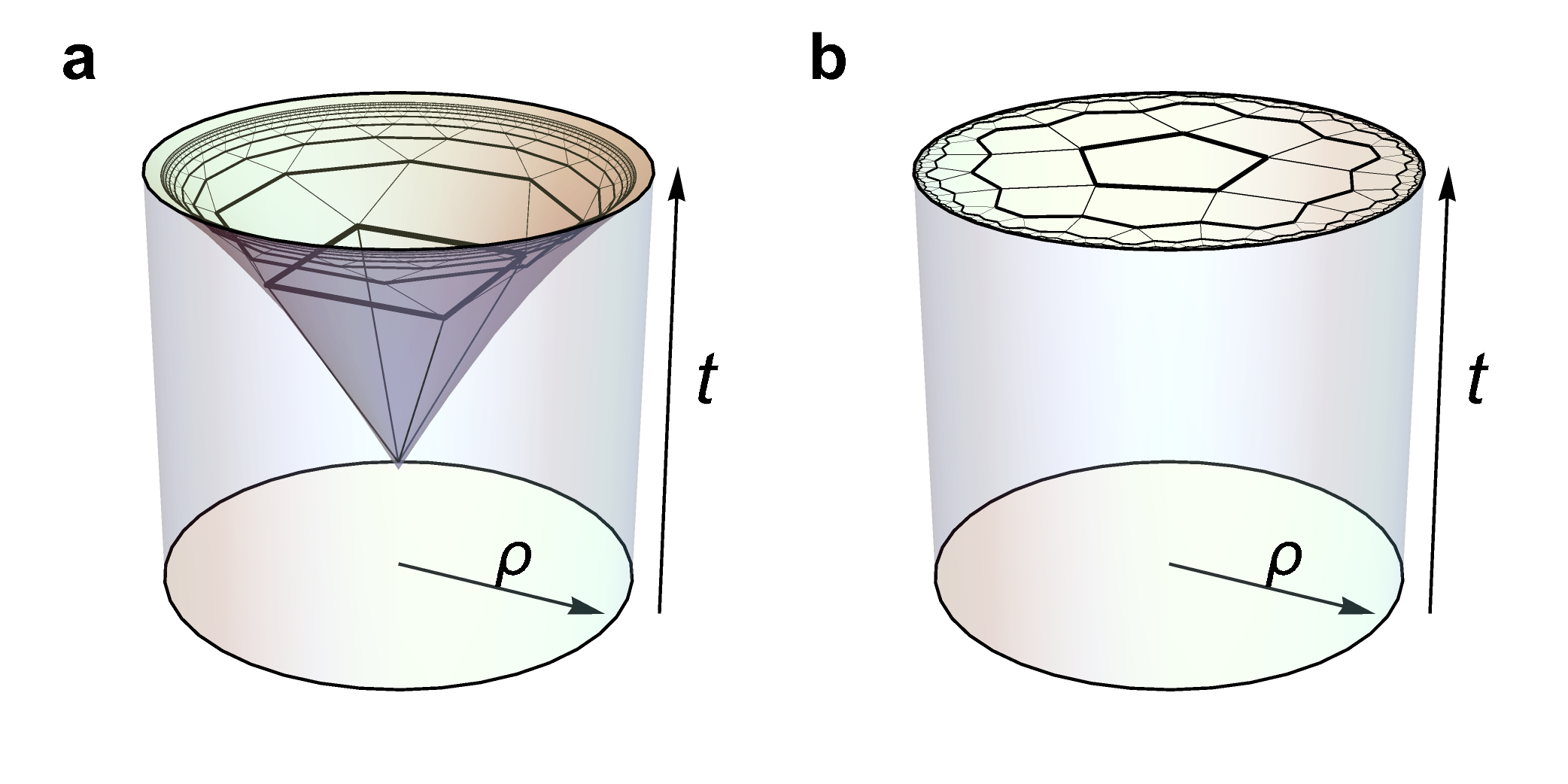}
\hspace{0.1cm}
\includegraphics[height=0.188\textheight]{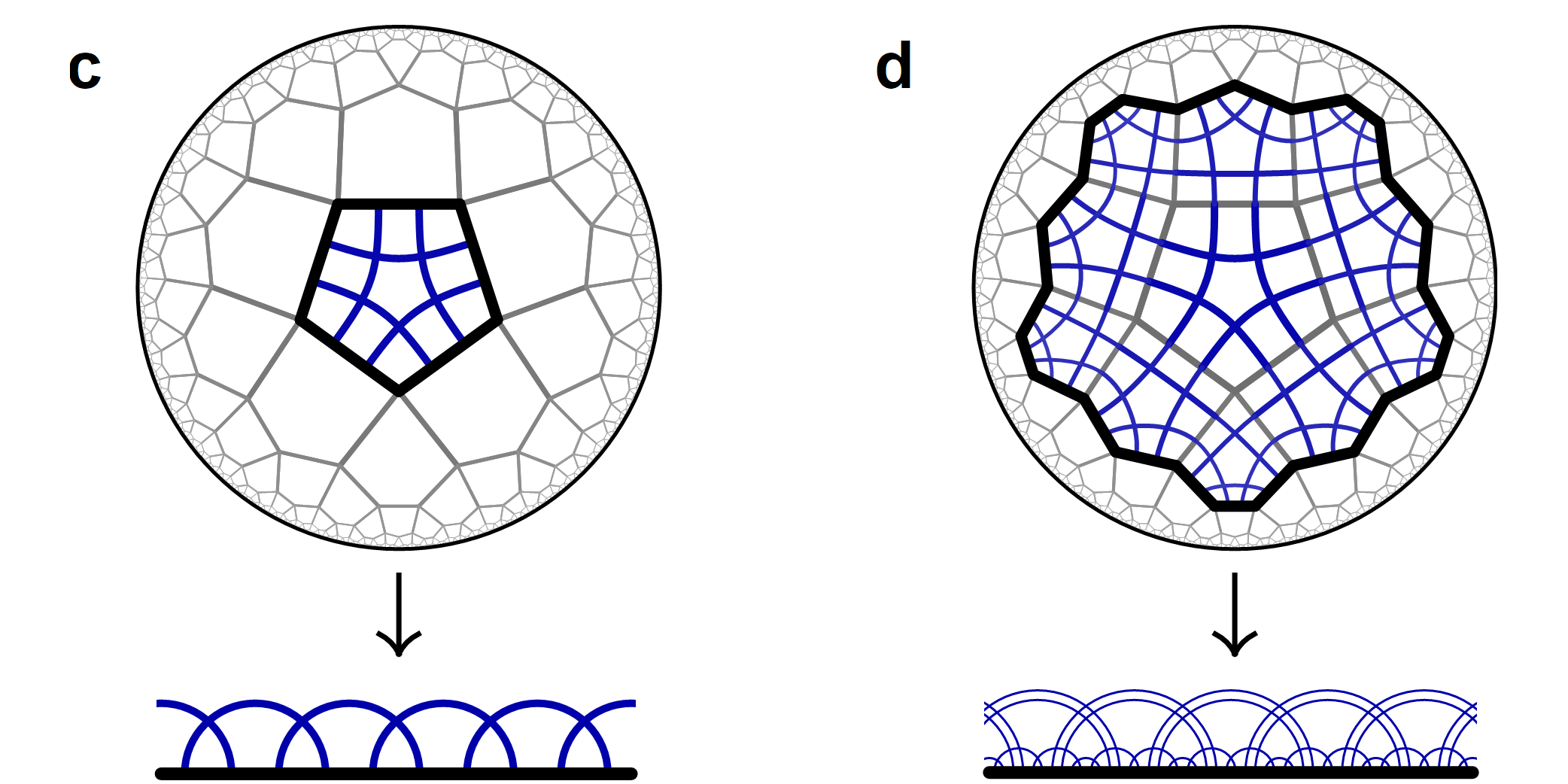}
\caption{\textbf{AdS embeddings and Majorana dimers.} \textbf{a-b,} embeddings of the MERA (a) and regular $\{5,4\}$ tiling (b) into AdS$_3$ space-time with coordinates $(t,\rho,\phi)$, with the AdS boundary located at the surface of the cylinder. The MERA corresponds to a light-cone embedding \cite{Milsted:2018san}, while regular discretizations such as the $\{5,4\}$ tiling can be identified with a time-slice. Both tilings can be divided into radial \emph{inflation layers}, which are drawn in bold.
\textbf{c-d,} the $\{5,4\}$ tiling with \emph{Majorana dimers} corresponding to boundary states of the hyperbolic pentagon code \cite{Jahn:2019nmz}. States at the zeroth and first inflation step are shown in the Poincar\'e disk (top) and half-plane projection (bottom).
}
\label{FIG_ADS_DISCRETIZATIONS}
\end{figure*}

\section{Results}
\subsection{Conformal symmetries and the Poincar\'e disk}
A central insight underlying AdS/CFT is that the $SO(d,2)$ symmetries of pure AdS space-time in $d{+}1$ dimensions are also the symmetries of a $d$-dimensional CFT, whose space-time can be identified with the asymptotic spatial boundary of the AdS bulk.
Breaking these continuous bulk symmetries by a discretization necessarily also breaks the boundary symmetries. 
Throughout this paper, we consider $2{+}1$-dimensional AdS$_3$ space-time, which can be projected onto cylinder coordinates $(t,\rho,\phi) \in (\mathbb{R},\, [0,1),\, [0,2\pi))$ with the metric
\begin{equation}
\label{EQ_ADS_PDISK}
\mathrm{d}s^2 = \frac{-(1+\rho^2)^2\, \mathrm{d}t^2 +  4 \alpha^2 \mathrm{d}\rho^2 + 4 \alpha^2 \rho^2 \mathrm{d}\phi^2}{(1-\rho^2)^2} \ ,
\end{equation}
where the constant $\alpha$ is the \emph{AdS radius} determining the global negative curvature of the geometry. 
The AdS boundary is located at $\rho \to 1$.
Instead of discretizing the entire space-time, we focus on space-like surfaces whose boundary is a slice of the AdS boundary at constant time $t$.
To describe the boundary quantum state on such a discretization, we consider a \emph{tensor network} embedded into the discretized surface: Each simplex of the discretization is associated with a tensor whose indices are contracted with indices of neighboring simplices\footnote{Note that this picture is the geometric dual of the Penrose graphical notation, where vertices represent tensors and connected edges contractions.}. The uncontracted indices are then located on the edges of the tiling's boundary, each index representing a site of a quantum state.
As shown in Fig.\ \ref{FIG_ADS_DISCRETIZATIONS}\textbf{a-b}, the bulk surface into which such a tensor network is embedded is not automatically a time-slice of AdS space-time. For example, the MERA can be identified with a path integral on a light-cone geometry \cite{Milsted:2018san}.
The geometry we consider here, however, is the AdS time-slice as shown in Fig.\ \ref{FIG_ADS_DISCRETIZATIONS}\textbf{b}, the same setup underlying the Ryu-Takayanagi formula \cite{PhysRevLett.96.181602} whose properties such tensor network constructions readily reproduce.
As with the MERA tensor network, these discretized time-slice geometries can be iteratively tiled, corresponding to a tensor network that is contracted in consecutive layers. Shown in Fig.\ \ref{FIG_ADS_DISCRETIZATIONS}\textbf{c-d} are the first two such contraction steps for an instance of Majorana dimer states, which can be contracted diagrammatically. We will use the dimer example in more detail below to demonstrate how the tiling symmetries are realized in tensor network states.

\begin{figure}[tb]
\centering
\includegraphics[width=0.4\textwidth]{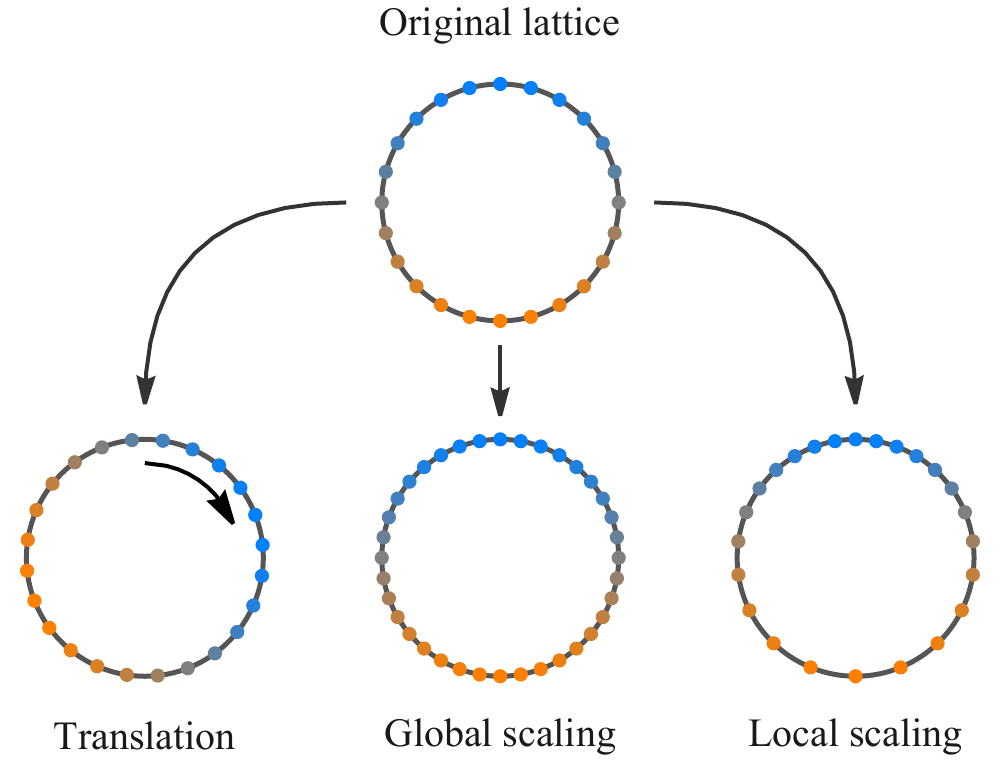}
\caption{\textbf{Invariance transformations of a CFT ground state.} The transformations are shown in their effect on a regular lattice with periodic boundary conditions and include translation of the lattice sites as well as global (uniform) and local (non-uniform) rescaling of the lattice. Here we consider local scaling transformations that preserve the total number of lattice sites.
}
\label{FIG_CONF_TRANSFORMATION}
\end{figure}

\begin{figure}[tb]
\centering
\includegraphics[width=0.48\textwidth]{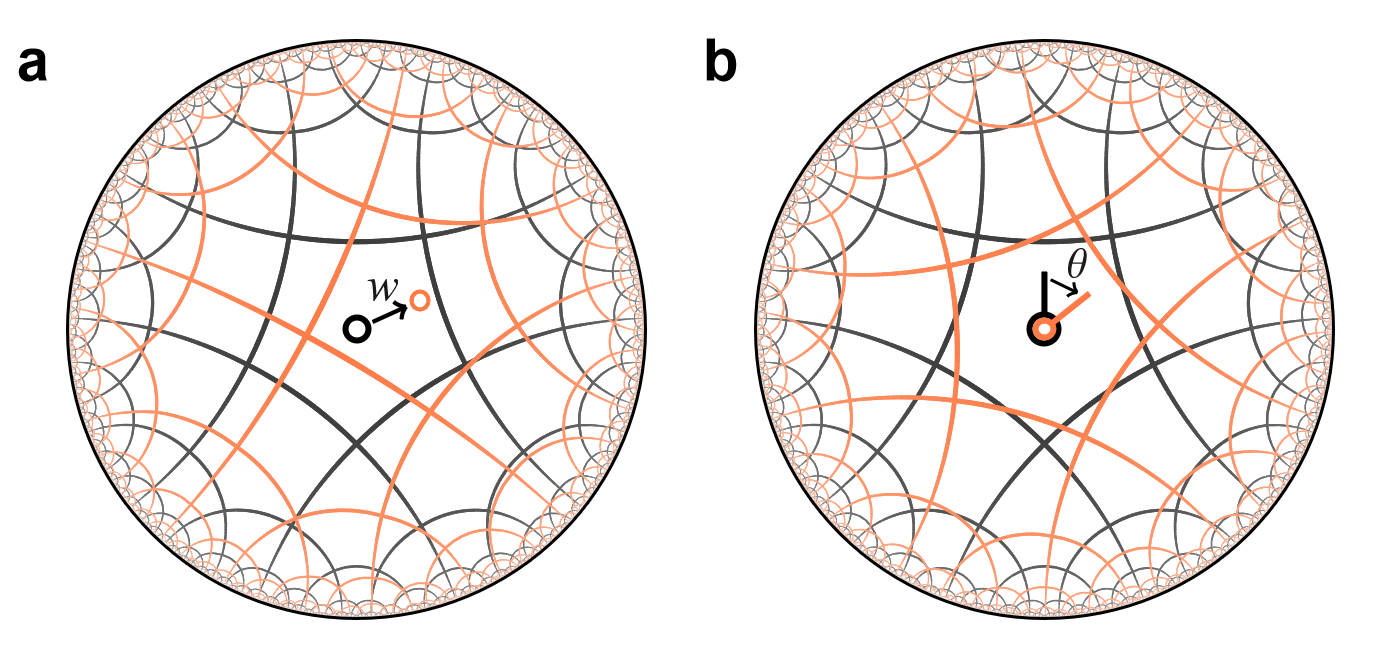}
\includegraphics[width=0.48\textwidth]{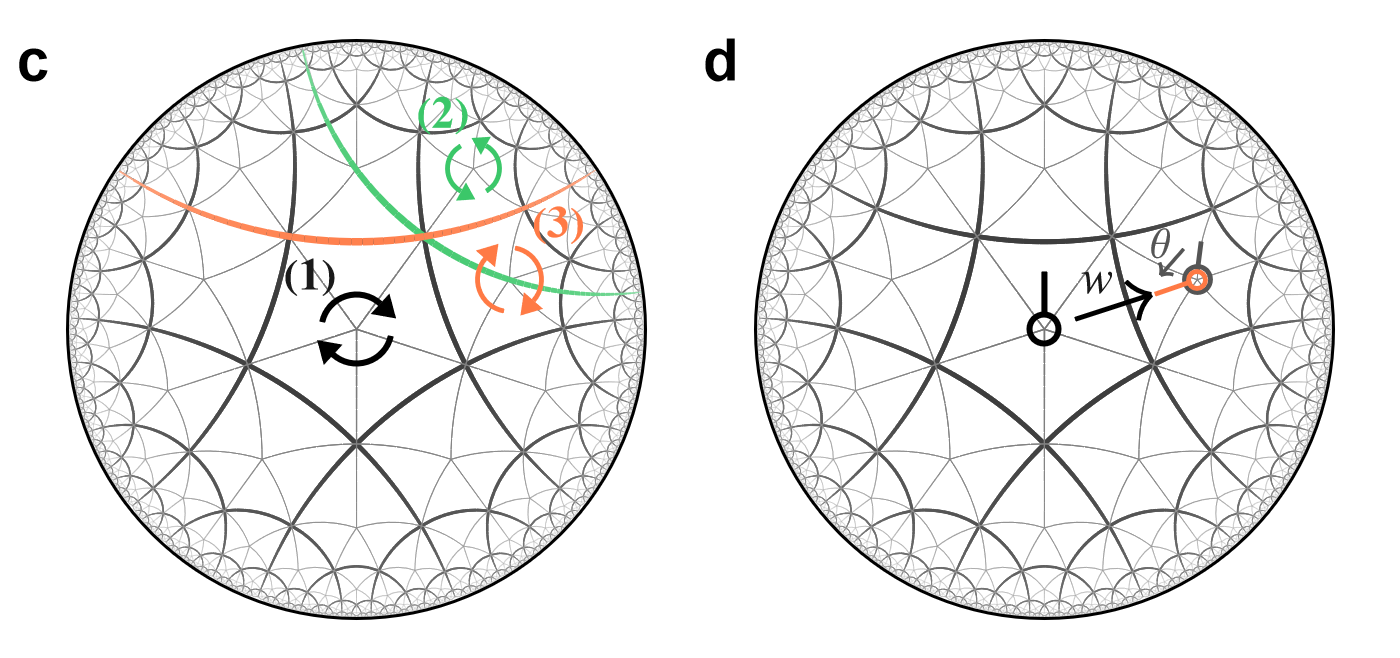}
\caption{\textbf{Symmetries of regular hyperbolic tilings. a-b,}
regular tilings of the Poincar\'e disk breaking the continuous $PSL(2,\mathbb{R})$ symmetry under transformations \eqref{EQ_BULK_SHIFT} that include 
\textbf{a,} bulk translations by $w$ not corresponding to a lattice shift and
\textbf{b,} bulk rotations that are not a multiple of $2\pi/k$.
\textbf{c-d,} triangle group transformations of triangulated tiling. An even number of reflections $(1)\mapsto(2)\mapsto(3)$ along triangle geodesics (\textbf{c}) is orientation-preserving (denoted by circular arrows on the central tile and its reflections) and equivalent to a bulk translation+rotation (\textbf{d}).
}
\label{FIG_CONF_BREAKING}
\end{figure}

An AdS time-slice in the continuum, equivalent to the \emph{Poincar\'e disk}, is invariant under $PSL(2,\mathbb{R})$ transformations. Describing a coordinate point by the complex number $z=\rho\, e^{\i \phi}$, these transformations take the form
\begin{equation}
\label{EQ_BULK_SHIFT}
z \mapsto z^\prime = e^{\i \theta} \frac{w + z}{1 + w^\star z} \,
\end{equation}
effectively shifting the origin of the Poincar\'e disk to the point $w$ with $|w|<1$ and rotating by an angle $\theta$. We refer to these as \emph{M\"obius transformations}, as $PSL(2,\mathbb{R})$ is a subgroup of the M\"obius group $PGL(2,\mathbb{C})$.
On the Poincar\'e disk boundary $\rho\to 1$, these transformations are equivalent to \emph{translations} and non-uniform \emph{local scale transformations}.
Furthermore, as length scales diverge at $\rho\to 1$, changing the cutoff scale $\rho_0$ corresponds to a uniform \emph{global scale transformation}.
These three transformations are exactly the spatial part of conformal transformations in $1{+}1$ dimensions, and are shown in Fig.\ \ref{FIG_CONF_TRANSFORMATION} in their action on a regular boundary lattice.
These local scale transformations act similarly to the spatial part of boost transformations in the AdS embedding spacetime, which also map between boundary regions of different size \cite{Casini:2011kv}.
Discretizing the bulk geometry restricts the symmetries on the boundary to a discrete subgroup of the previously continuous symmetries. As in the continuous case, we can characterize these boundary symmetries by the group of invariance transformations in the bulk, which are characterized by a number of symmetry groups. Specifically, we consider regular $\{n,k\}$ tilings (using the \emph{Schl\"afli notation}) that are composed of regular $n$-gons, $k$ of which are adjacent at each corner. Symmetry groups of physical models on these tilings have previously been identified and discussed in the context of tensor networks \cite{Bhattacharyya:2016hbx,Osborne:2017woa,Kohler:2018kqk} as well as physical bulk models and their band theory \cite{Maciejko:2020rad,Boettcher:2021njg}.
In any regular tiling -- hyperbolic or not -- the reflection along edges of the tiling leaves the tiling invariant, and general mappings of the tiling onto itself can be composed from such reflections, defining the symmetries of a \emph{Coxeter group}.
As any polygon tiling can be further subdivided into triangles, the symmetries of regular tilings extend to those of a hyperbolic \emph{triangle group}, i.e., the group of reflections along edges in a triangular tiling.
Note that in the case of flat and hyperbolic tilings, these are infinite symmetry groups assuming the tiling to be infinitely extended; cutting it off after a finite number of tiles inherently breaks these symmetries.
For physical symmetry transformations that preserve orientation, we restrict ourselves to a subgroup of index two of the triangle group consisting of all transformations composed of an even number of reflections, called \emph{von Dyck groups}. 
In the hyperbolic case, these transformations are now discrete subgroups of the continuous $PSL(2,\mathbb{R})$ transformations, as Fig.\ \ref{FIG_CONF_BREAKING}\textbf{c-d} demonstrates.
This property defines a \emph{Fuchsian group}.
Let us consider these symmetries in explicit examples. 
As we see in Fig.\ \ref{FIG_CONF_BREAKING}\textbf{a-b}, a M\"obius transformation \eqref{EQ_BULK_SHIFT}, such as a bulk translation of the tiling center to a point $w$ (\ref{FIG_CONF_BREAKING}\textbf{a}) or a rotation by an angle $\theta$ (\ref{FIG_CONF_BREAKING}\textbf{b}), generally does not map a regular hyperbolic tiling of infinite size onto itself.
The only tiling-preserving transformations which also preserve orientation consist of bulk translations to a point $w$ that is the center of an $n$-gon tile (equivalently, a lattice vector in the dual $\{k,n\}$ tiling), followed by a suitable rotation with an angle $\theta$ that re-aligns the tiling up to a $\mathbb{Z}_n$ ambiguity.
Such a transformation is equivalent to an even number of reflections along a geodesic edge of the triangularized tiling, demonstrated in Fig.\ \ref{FIG_CONF_BREAKING}\textbf{c-d}, which is exactly the construction of Fuchsian groups via hyperbolic von Dyck groups mentioned above.
In order to understand these bulk tiling transformations as discretizations of conformal transformations on the boundary, we need to study their effect for a finite tiling cutoff.
We will now discuss the construction of such a cutoff in general and in a specific model before returning to these specific transformations.

\subsection{Global scaling and inflation rules}

\begin{figure}[t]
\centering
\includegraphics[width=0.52\textwidth]{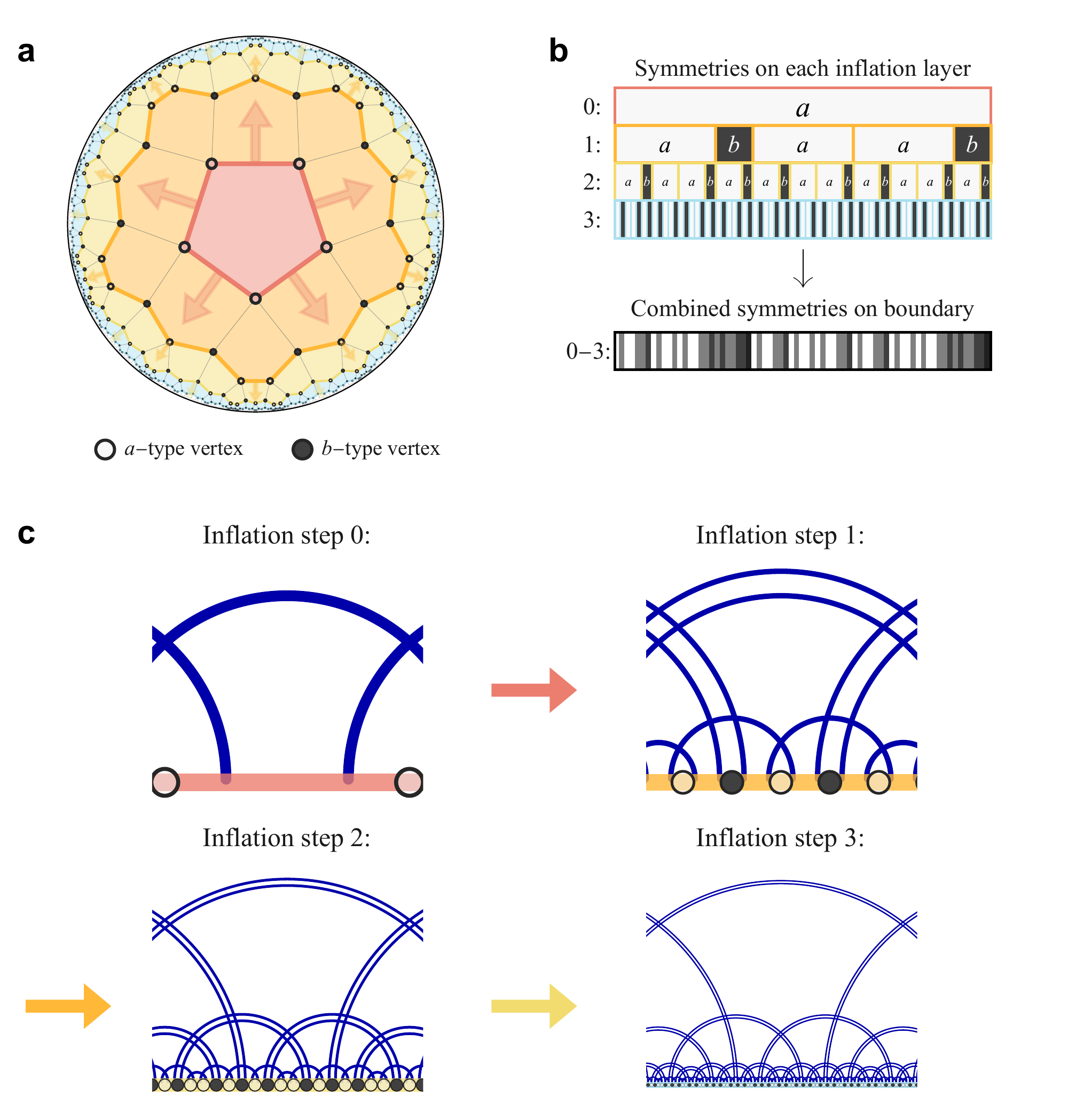}
\caption{\textbf{Global scale transformations. a,} vertex inflation of the hyperbolic $\{5,4\}$ tiling with $a$- and $b$-type boundary vertices following the rule \eqref{EQ_VINFL_54}, with an overall $\mathbb{Z}_5$ symmetry.
\textbf{b,} the multi-scale quasicrystal ansatz (MQA), a visualization of layered quasiperiodic symmetries. Its layers consist of sequences of vertex types on one fifth of the total system (one $a$ letter on first layer), the ``product'' of whose quasiperiodic distribution (bottom row) determines the symmetries of the boundary state. 
\textbf{c,} corresponding Majorana dimer state from a single initial edge.
Note the asymptotic pairing of dimers as discussed in Ref.\ \cite{Jahn:2019nmz}.
}
\label{FIG_GLOBAL_SCALING}
\end{figure}

A size cutoff for regular hyperbolic tilings is usually defined via an iterative tiling construction that starts with a central tile.
Equivalently, a tensor network on such a geometry can be iteratively contracted, producing a boundary state of increasingly large but finite Hilbert space dimension after each iteration step.
This fine-graining procedure can be described more precisely by specifying the \emph{inflation rule} under which the bulk tiling is filled layer-by-layer, or \emph{inflated} \cite{Boyle:2018uiv}.
One choice for such an inflation rule is \emph{vertex inflation} (also called \emph{vertex completion}), where each inflation step adds a closed layer of tiles adjacent to the vertices of the previous tiling. This rule is a particularly natural one, as its behavior is preserved under a mapping from an $\{n,k\}$ tiling to its dual $\{k,n\}$ one. A vertex inflation step also underlies the construction of \emph{hyperinvariant tensor networks} \cite{PhysRevLett.119.141602} that realize certain properties of the MERA in specific tensor networks on regular hyperbolic tilings.
Such an inflation method acts as a discrete global scale transformation on the boundary states of a tensor network: Each inflation step increases the number of boundary sites by a scale factor $\lambda$ that approaches a constant independent of the initial starting region. While restricted to $\lambda\in\mathbb{N}_{>1}$ for the MERA, this factor can take a large range of fractional values for an $\{n,k\}$ tiling. Its asymptotical value after many vertex inflation steps is given analytically by \cite{Jahn:2019mbb}
\begin{equation}
\lambda = \frac{2 + f(n,k) + \sqrt{(4 + f(n,k))f(n,k)}}{2} \ ,
\end{equation}
where $f(n,k) = n k - 2(n+k)$ is positive for any hyperbolic $\{n,k\}$ tiling.
The geometrical structure of the boundary layers is not asymptotically smooth but \emph{quasiperiodic}, with self-similarity on all scales \cite{Boyle:2018uiv}.
As a consequence, the scaling factor $\lambda$ not only describes the growth of the total system size but also that of any sufficiently fine-grained boundary region.
Specific vertex inflation rules can be written in terms of replacement rules on letter sequences, with each letter denoting specific boundary features \cite{Boyle:2018uiv}. 
For the $\{5,4\}$ tiling, we follow the notation of Ref.\ \cite{Jahn:2019mbb} and denote with $a$ and $b$ the types of vertices that appear on tiling boundaries throughout vertex inflation, with each vertex either connected to a single boundary tile (type $a$) or sitting between two of them (type $b$).
A vertex inflation step then corresponds to applying the rule
\begin{align}
\label{EQ_VINFL_54}
a &\mapsto abaab \ , &
b &\mapsto ab \ , 
\end{align}
where the strings correspond to letter sequences. 
Fig.\ \ref{FIG_GLOBAL_SCALING}\textbf{a} shows the vertex inflation layers of the $\{5,4\}$ tiling as colored bands, with the boundary vertices on each layer denoted as circles and dots depending on their type.
For this tiling, the starting configuration at the $0$th inflation step corresponds to the sequence $aaaaa$, resulting in a five-fold $\mathbb{Z}_5$ symmetry of the layers around the central pentagon.
Rearranging a fifth of the letter sequence at each layer (i.e., inflation steps for a single letter $a$) into linear blocks leads to the visualization in Fig.\ \ref{FIG_GLOBAL_SCALING}\textbf{b}.
We refer to this approach of stacking the quasiperiodic symmetries on each inflation layer as the \emph{multi-scale quasicrystal ansatz}, 
which serves as a simplified model of tensor networks on regular hyperbolic lattices grown under vertex inflation.
As shown in Ref.\ \cite{Boyle:2018uiv}, the letter sequence at the $n$th inflation step fully describes the boundary geometry at that step, but boundary states produced by tensor networks on such a geometry should be expected to have more complicated symmetries, with contributions from all layers relevant for the final state. As tensor networks on hyperbolic  tilings typically produce \emph{critical} states with divergent correlation lengths (see Ref.\ \cite{Jahn:2017tls} for the Gaussian fermionic case), the symmetries on all, and not merely short length scales, contribute to the final result.
This is sketched in the bottom row of Fig.\ \ref{FIG_GLOBAL_SCALING}\textbf{b}, where we visualize the final state symmetries as a local multi-scale ``product'' of contributions from each layer of the MQA. While the exact form of this product depends on the tensor content, the multi-scale quasiperiodic structure is fixed by the geometry.

\subsection{Scale transformation with dimers}

To demonstrate scale transformations on boundary states of a specific tensor network, we consider tensors described by \emph{Majorana dimer states}: These are Gaussian fermionic states, composed of paired Majorana modes, that are characterized by efficiently contractible tensors and can also be interpreted as ground states of Gaussian \emph{stabilizer Hamiltonians} in the context of quantum error correction. Its most relevant example for our purposes is the $[[5,1,3]]$ code that encodes one logical qubit in five physical spins or fermions and can correct one Pauli-type error \cite{Bennett:1996gf,Laflamme:1996iw}. The logical code states $\bar{0}$ and $\bar{1}$ are encoded as Majorana dimer states on the physical fermions. These share the same correlation pattern and can be visualized as
\begin{align}
\label{EQ_HYPEC_STATES}
\ket{\bar{0}}_5\; &= \;
\begin{gathered}
\includegraphics[height=0.08\textheight]{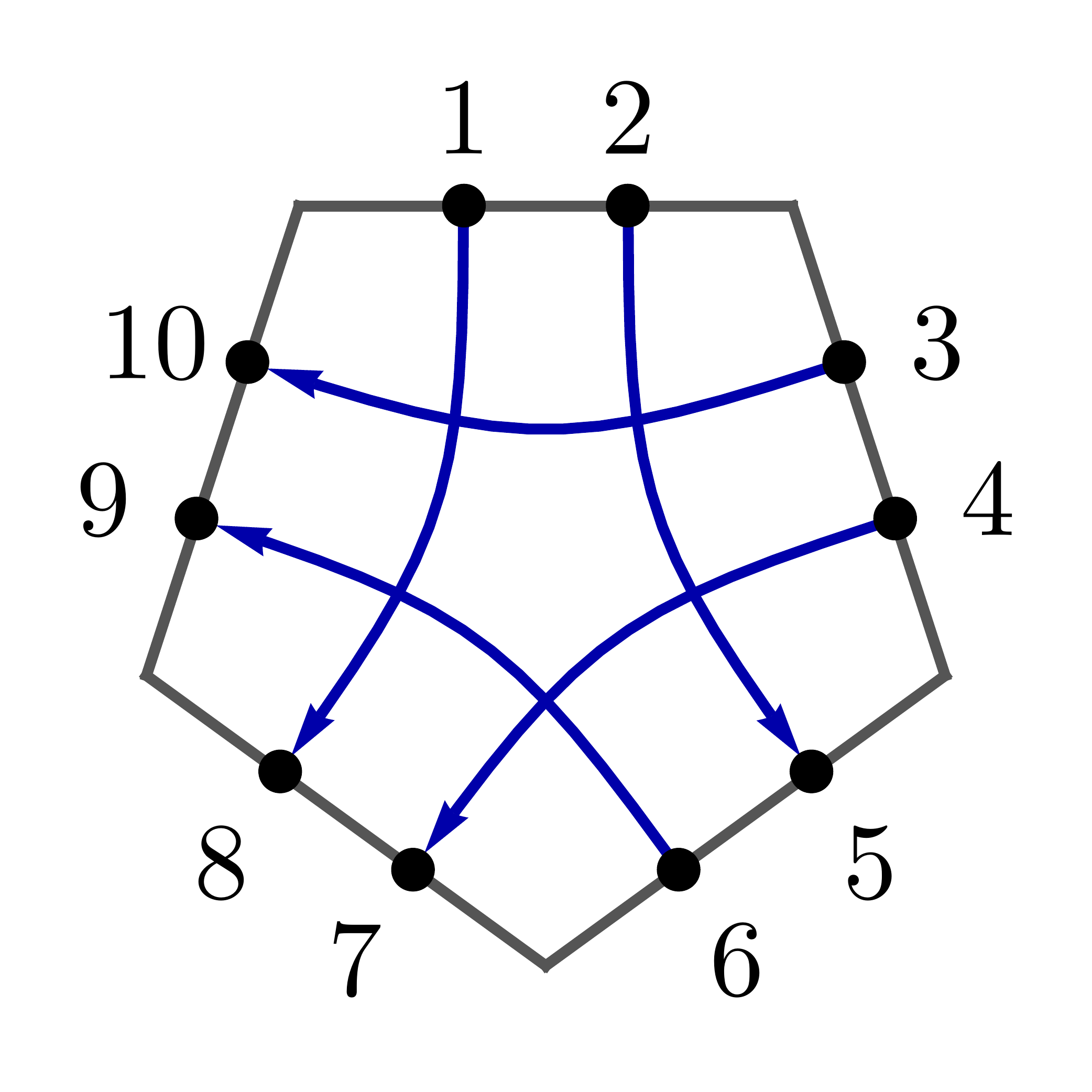}
\end{gathered} \ , &
\ket{\bar{1}}_5\; &= \;
\begin{gathered}
\includegraphics[height=0.08\textheight]{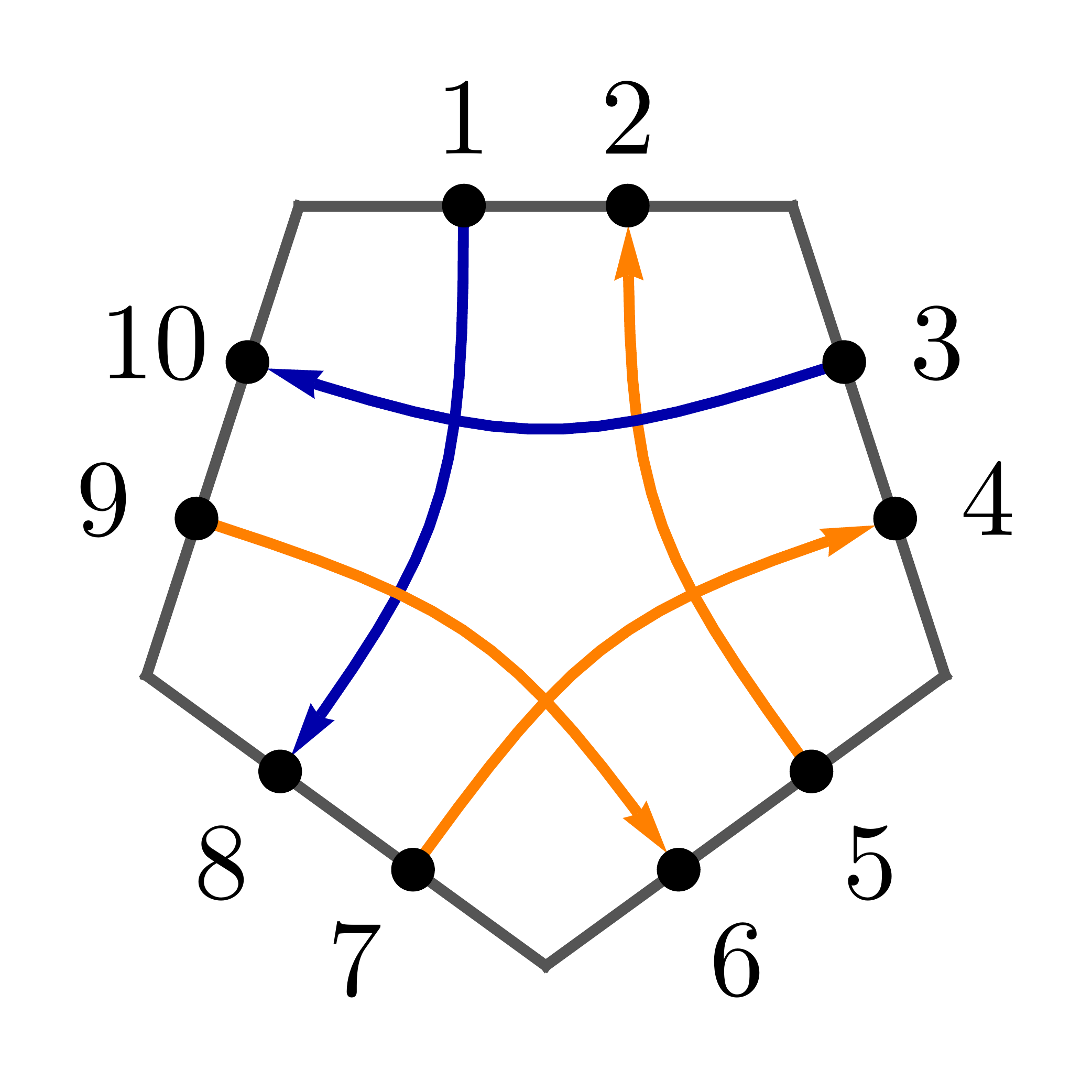}
\end{gathered} \ .
\end{align}
A pentagon in this visualization represents a state on five qubits, or equivalently, a five-leg tensor encoding this state.
Each arrow from Majorana mode $j$ to $k$ in this diagram corresponds to a term $\i\m_j\m_k$ in the stabilizer Hamiltonian while also representing a maximally entangled pair of Majorana modes.
We can use these states as the building blocks of a class of holographic models known as \emph{HaPPY codes} \cite{Pastawski2015} in which these code states are placed on a regular hyperbolic tiling. When doing so, the contraction of these states (or rather, the tensors encoding them) into a full tensor network is equivalent to fusing endpoints of Majorana dimers between individual diagrams, turning the contraction process into a geometric problem of tracing connected dimers throughout the network. 
As a result, large tensor contractions can be performed analytically. In particular, entanglement entropies and correlation functions can be computed for various hyperbolic geometries, leading to results consistent with those of conformal systems \cite{Jahn:2019nmz}.
Given a tiling cutoff, tensor contraction of the $[[5,1,3]]$ code states in Majorana dimer form as shown in Fig.\ \ref{FIG_ADS_DISCRETIZATIONS}\textbf{c-d} thus requires connecting all dimers on tiles up to the cutoff. Moving the cutoff closer to the Poincar\'e disk boundary leads to a boundary dimer state supported on more modes, successively fine-graining the resulting correlation structure \cite{Jahn:2019nmz,Jahn:2019mbb}.
In the vertex inflation setup, this is shown in Fig.\ \ref{FIG_GLOBAL_SCALING}\textbf{c} for four inflation steps. The dimer patterns follow the same sequence of $a,b$ letters as the boundary geometry, with an $a$ letter corresponding to a local crossing of a pairs of dimers, while dimers around a $b$-type vertex form non-crossing singlets. 
The dimer representation of the boundary state supports the qualitative picture of the MQA from Fig.\ \ref{FIG_GLOBAL_SCALING}\textbf{b}: Each inflation step adds new (dimer) degrees of freedom at small scales while preserving the long-range correlation structure, resulting in a boundary state with symmetry contributions from all length scales.
The inflation steps are renormalization group (RG) steps not only in terms of correlations, as we have just seen, but can be also be seen as fine-graining entanglement:
Along with the scale factor $\lambda_{\{5,4\}} = 2 + \sqrt{3}$, one first computes the asymptotic scaling of the length of any discrete bulk geodesic $|\gamma_A|$ through each inflation step.
Consider a tensor network of bond dimension $\chi$.
For a minimal cut $\gamma_A$ through this network with the same endpoints as a boundary region $A$, its entanglement entropy \cite{AreaReview} is upper-bounded as
\begin{equation}
\label{EQ_RT_BOUND}
S_A \leq \frac{|\gamma_A|}{s} \log \chi \ ,
\end{equation}
where $s$ is the geodesic length of each edge, which is constant for regular tilings.
One can bound the scaling of $S_A$, and by extension compute the maximal central charge $c^\text{max}$ \cite{CalabreseReview} of the boundary state, by calculating the asymptotic growth of $|\gamma_A|$ for a given $\{n,k\}$ tiling \cite{Jahn:2019mbb}.
Furthermore, it is possible to relate $c^\text{max}$ to the Gaussian curvature $K=-1/\alpha^2$ of the embedding Poincar\'e disk. The AdS radius $\alpha$ is then given by
\begin{equation}
\label{EQ_ADS_GEODESIC}
\frac{s}{\alpha} = 2 \arcosh\left( \frac{\cos\frac{\pi}{n}}{\sin\frac{\pi}{k}} \right) = 2\log\left(\frac{2k}{\pi} \cos\frac{\pi}{n} \right) + O(k^{-2}) \ .
\end{equation}
Combining \eqref{EQ_RT_BOUND} and \eqref{EQ_ADS_GEODESIC} leads to a discrete, tiling-dependent generalization of the Brown-Henneaux formula $c = 3\alpha/(2 G)$ \cite{Brown:1986nw} relating AdS radius $\alpha$ and central charge $c$. 
In this identification the effective gravitational constant scales as $G~\sim~1/\log\chi$; comparing with AdS/CFT, the limit of semiclassical bulk gravity thus corresponds to a model with large bond dimension. Geometrically, this can be represented as a flat subdivision of each tile into smaller tiles, each of which carries only spin degrees of freedom on its edges. 
This discrete version of the Brown-Henneaux formula is generally an upper bound $c<c^\text{max}$ that can only be saturated for tensor networks that are maximally entangling across any minimal cut.
However, given the specific tensors of a tensor network on a regular hyperbolic geometry, we can analytically compute the value of the effective central charge $c$ using the same approach.
Consider the HaPPY code in its Majorana dimer formulation in Figs.\ \ref{FIG_ADS_DISCRETIZATIONS}\textbf{c-d} and \ref{FIG_GLOBAL_SCALING}\textbf{c}: 
As each dimer carries an entanglement entropy of $\frac{1}{2} \log 2$, an inflation step is equivalent to an \emph{entanglement renormalization} step \cite{Vidal.ERintro}, adding entanglement through the addition of dimers along the boundary.
Relating the growth of boundary regions to the growth of entanglement over cuts under an inflation step allows us to analytically compute $c$. For example, for the $\{5,4\}$ HaPPY model under vertex inflation one finds \cite{Jahn:2019mbb}
\begin{align}
\label{EQ_C_54}
c_{\{5,4\}} &= \frac{9 \log 2}{\log\left(2 + \sqrt{3}\right)} \approx 4.74 \ .
\end{align}
As this is a bond dimension $\chi = 2$ model with only spin degrees of freedom on the edges of each tile, we find a relatively small central charge. For a more realistic model of (semiclassical) bulk gravity, a large bond dimension would be required, e.g.\ by placing  $N$ copies of the $[[5,1,3]]$ code tensor on each tile, increasing $c$ by a factor of $N$.
Note that the procedure leading to \eqref{EQ_C_54} can be applied to any tensor network on a regular hyperbolic tiling whose tensors have equal entanglement properties in the asymptotic region near the Poincar\'e boundary, yielding a different central charge for each model.
A simpler version of this approach can also be used to compute the change in qubit erasure probabilities with each inflation step: While the size of boundary regions grows exponentially and the entanglement entropy linearly with the number of inflation steps, the erasure probability of a logical qubit deep in the bulk decays double exponentially if the single-qubit erasure probability is below a threshold value \cite{Pastawski2015}. Like the effective central charge, this threshold is lattice-dependent.

\subsection{Local scaling}

\begin{figure*}[t]
\centering
\includegraphics[width=0.95\textwidth]{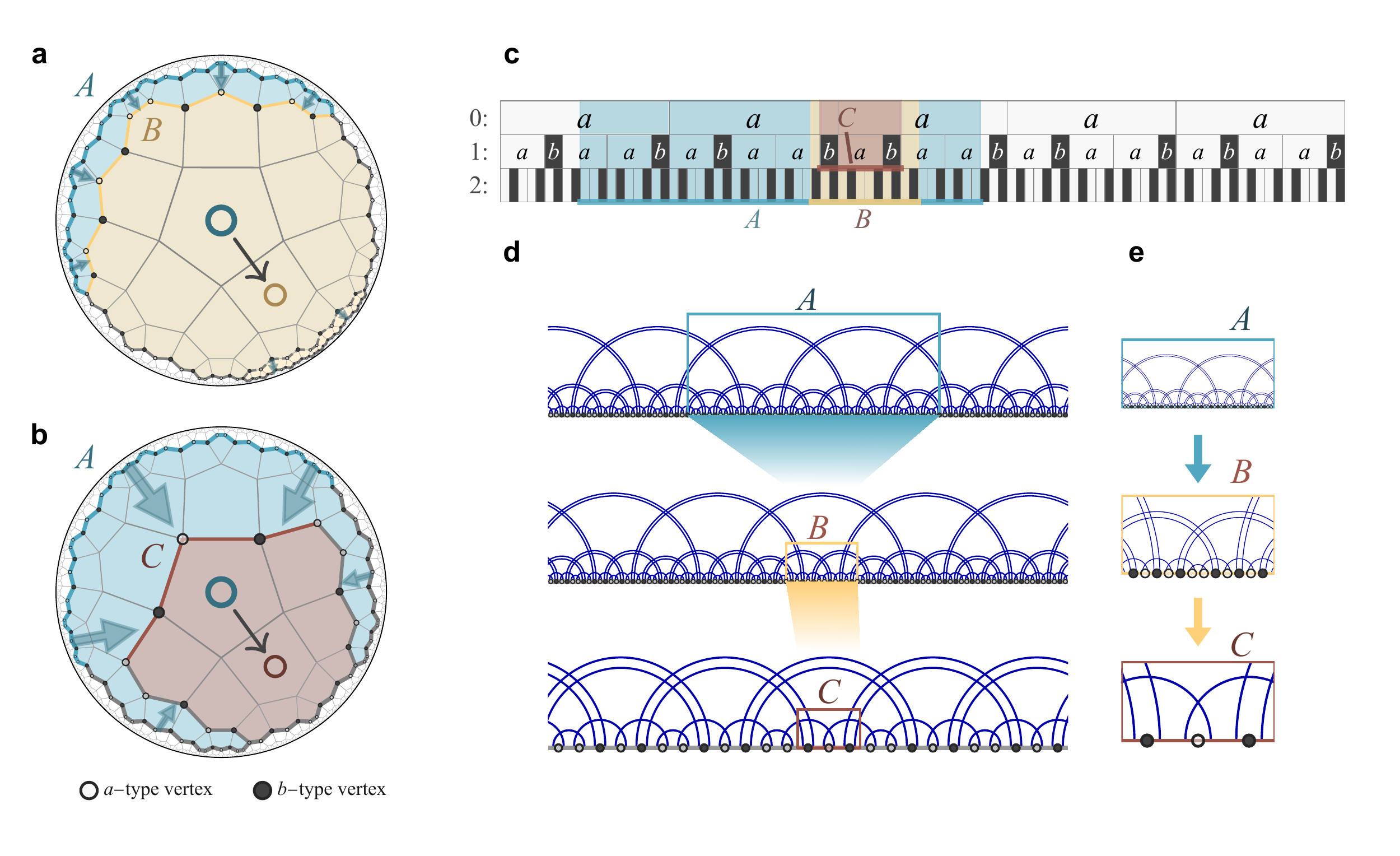}
\caption{\textbf{Local scale transformations. a,} bulk shift of $\{5,4\}$ tiling center (thick circle) after two inflation steps, coarse-graining boundary region $A$ (blue) to $B$ (yellow) on upper-left part of the tiling, as well as fine-graining a corresponding region on the lower-right one.
\textbf{b,} the same bulk shift with an additional deflation step, coarse-graining boundary region $A$ to $C$ while leaving the lower-right part of the tiling unchanged.
\textbf{c,} the regions $A,B,C$ from \textbf{a-b} in the MQA of Fig.\ \ref{FIG_GLOBAL_SCALING}\textbf{b}, with each row representing an inflation layer. Note that region $C$ corresponds to only two inflation layers.
\textbf{d,} the total HaPPY Majorana dimer state after two (first two rows) and one inflation step (bottom row), with subsystems $A,B,C$ marked.
\textbf{e,} the subsystems $A,B,C$ of the HaPPY Majorana dimer state rescaled to the same size for comparison.
The same correlation structure appears on multiple length scales in different subsystems of the state, with only small deviations.
}
\label{FIG_LOCAL_SCALING}
\end{figure*}

The distinguishing feature of conformal transformations is a local rather than merely global change in scale, i.e., a location-dependent fine- and coarse-graining. These transformations can be represented by discrete lattices, such as for the MERA tensor network \cite{Czech:2015xna}.
Regular tilings also naturally incorporate local scale transformations in the shape of non-uniform boundary deformations that arise from the discretization of the bulk M\"obius transformations \eqref{EQ_BULK_SHIFT}. 
As in the continuum, the corresponding bulk transformation is a lattice-preserving shift $w$ of the center of the Poincar\'e disk combined with a global rotation $\theta$ as shown in Fig.\ \ref{FIG_CONF_BREAKING}.
We show in Fig.\ \ref{FIG_LOCAL_SCALING}\textbf{a} how such a bulk shift on a tiling after two inflation steps moves parts of the tiling boundary while changing its resolution: While one region of the boundary is fine-grained in a local application of the vertex inflation rules, another region $A$ is coarse-grained to a region $B$ with fewer edges through local vertex \emph{deflation} rules. The boundary geometry as a whole is left invariant by this transformation, as each vertex in the hyperbolic tiling can be arbitrarily defined as its ``center''. In other words, the bulk shift only changes the projection of the tiling onto the Poincar\'e disk.
Note that the effective local scale transformation on the boundary, following the same inflation/deflation rules as the previously described global scale transformation, thus exhibits the same scaling factor $\lambda$ for large regions. 
Unlike arbitrary deformations of the tiling boundary, these transformations thus preserve the entanglement scaling and effective central charge $c$ computed above for global scale transformations.
Combining the bulk shift with a deflation step, as Fig.\ \ref{FIG_LOCAL_SCALING}\textbf{b} shows, preserves the (previously inflated) part of the boundary while deflating the rest of the boundary. 
Again using the $\{5,4\}$ HaPPY code in its Majorana dimer form for illustration, Fig.\ \ref{FIG_LOCAL_SCALING}\textbf{d} shows how the local deflation step acts on the boundary state while leaving it globally invariant.
In Fig.\ \ref{FIG_LOCAL_SCALING}\textbf{e} these subsystems are rescaled to the same size to show that the local correlation structure is preserved (up to coarse-graining).
As there are as many such allowed transformations as there are tiles at a given inflation step, these local scale transformations lead to a fractal self-similarity of the boundary state as the number of inflation steps is increased. 
We can also characterize these transformations using the MQA picture introduced above, as is visualized in Fig.\ \ref{FIG_LOCAL_SCALING}\textbf{c}. The two examples from Fig.\ \ref{FIG_LOCAL_SCALING}\textbf{a-b} correspond to a mapping from a large block $A$ of three quasiperiodic layers to a smaller block $B$ contained within it (\textbf{a}) and to a small block $C$ of only two layers (\textbf{b}).
Close inspection shows that all of the three blocks can be traced to similar initial letter sequences on different layers: For example, the sequence $bab$ belonging to the red-shaded block $C$ in Fig.\ \ref{FIG_LOCAL_SCALING}\textbf{c} is inflated to 
\begin{equation}
bab \mapsto ababaabab
\end{equation}
which belongs to the yellow-shaded block $B$, but also appears as a subsequence under inflation of another sequence,
\begin{equation}
aaa \mapsto aba\underline{ababaabab}aab \ .
\end{equation}
This sequence is indeed the starting point of the blue-shaded (initial) block $A$ of both transformations considered in Fig.\ \ref{FIG_LOCAL_SCALING}, showing how identical sequences appear on different scales as a consequence of an invariance under discretized M\"obius transformations.
The number of such transformations that map the boundary system to part of itself is given by the number of tiles up to the given inflation step (each defining a possible bulk shift) times the number of inflation steps (possible rescalings). The former is propertional to the number of boundary sites and the latter to the discretized length scales encoded, as would be expected from a discretization of local scale transformations.
This self-similarity can also be explained by the MQA, where we find that a given letter (say, $b$) on any inflation level produces the same sequences ``below'' it and thus a similar correlation structure in the corresponding boundary subsystem up to boundary effects. 
Of course, the block-based MQA as shown in Figs.\ \ref{FIG_GLOBAL_SCALING}\textbf{b} and \ref{FIG_LOCAL_SCALING}\textbf{c} is dependent on the exact assignment of the inflation rules and does not take into account any ``smearing'' of correlations that a tensor network with less localized correlations than Majorana dimers would produce. However, the self-similarity argument is indendendent of the letter assignment (e.g.\ $a \,{\mapsto}\, abaab,b \,{\mapsto}\, ab$ instead of $a \,{\mapsto}\, baaba,b \,{\mapsto}\, ba$) and self-similarity of subsystems is still expected for subsystem sizes larger than the ``smearing radius'' by which the tensors in the particular tensor network spread correlations.
This self-similarity, as in continuum CFTs, enforces a polynomial decay of correlation functions, as short- and long-range correlations are equivalent up to a factor corresponding to the scaling dimension of the fields considered. We elaborate on the connection to continuum CFTs below.

\subsection{Approximate translation invariance}

\begin{figure*}[t]
\centering
\includegraphics[width=0.95\textwidth]{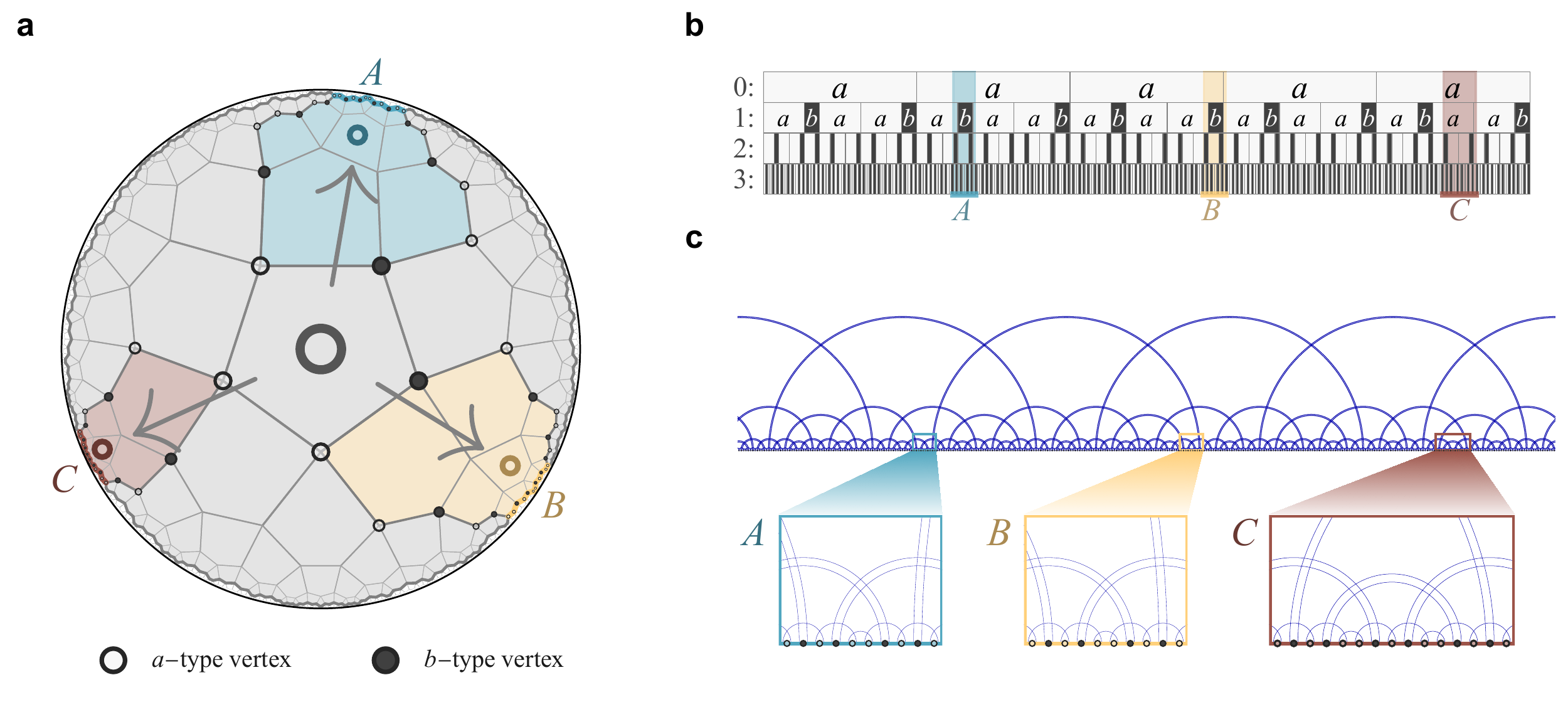}
\caption{\textbf{Boundary translations. a,} the $\{5,4\}$ tiling after three inflation steps (gray-shaded area) with the tiling center (thick circle) on the central tile.
Colored areas correspond to a bulk shift of tiling center and two deflation steps (similar to Fig.\ \ref{FIG_LOCAL_SCALING}\textbf{b}) so that boundary regions $A,B,C$ of the smaller sets of tiles overlap with the boundary of the large (gray-shaded) one.
\textbf{b,} the regions $A,B,C$ from \textbf{a} in the MQA picture of Fig.\ \ref{FIG_GLOBAL_SCALING}\textbf{b}, with each row representing an inflation layer.
\textbf{c,} the total HaPPY Majorana dimer state after three inflation steps, with the subsystems $A,B,C$ magnified. All three show a similar correlation structure and quasiperiodic boundary sequences, as they can be identified with parts of the same deflated geometry related by bulk shifts in \textbf{a}.
}
\label{FIG_TRANSL_INV}
\end{figure*}

\begin{figure}[tb]
\centering
\includegraphics[width=0.5\textwidth]{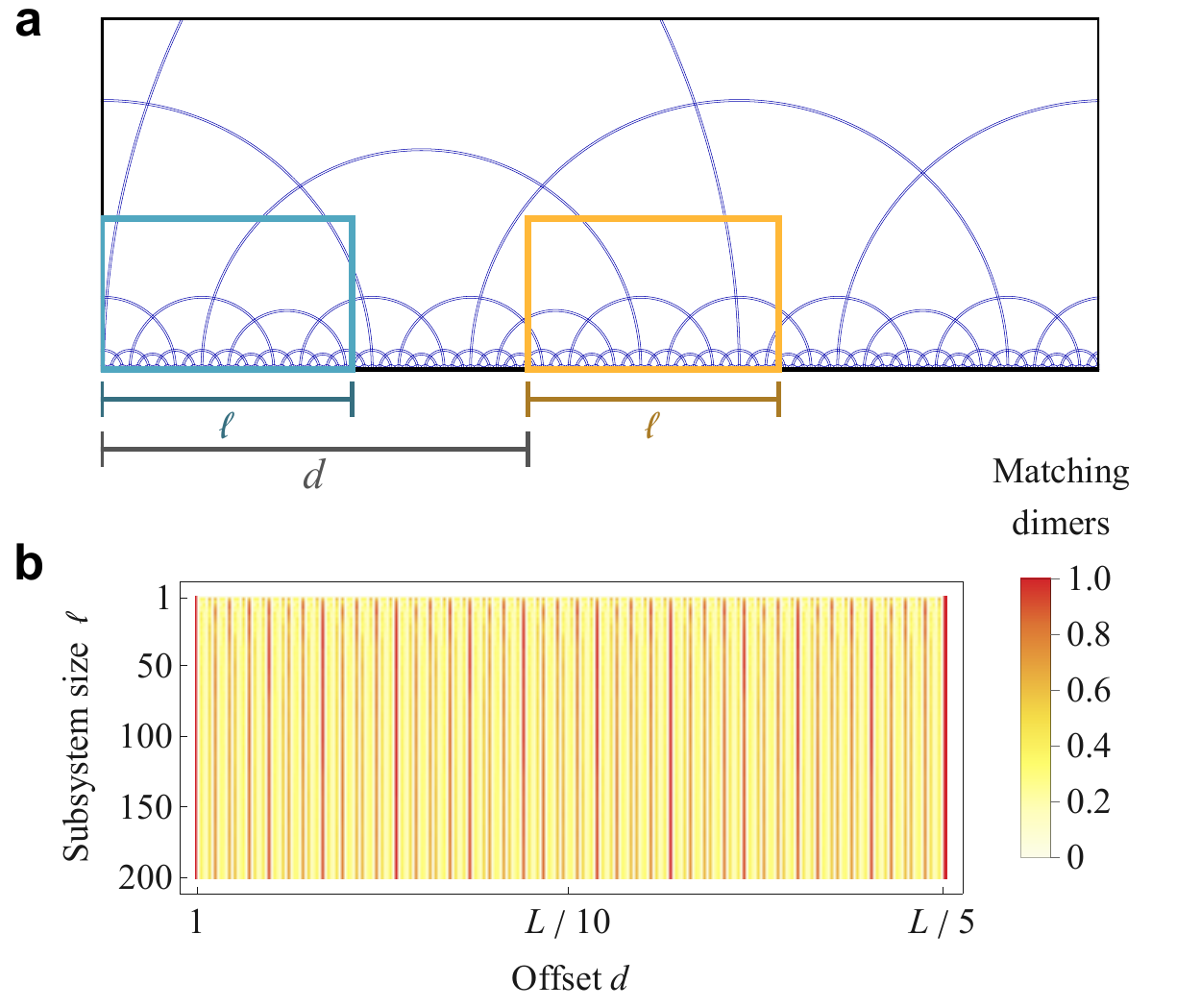}
\caption{\textbf{Approximate translation invariance of subsystems of HaPPY code dimers. a,} 
two dimer subsystem of $\ell$ sites ($2\ell$ Majorana modes) with a relative offset of $d$ sites for a total system of $L=1325$ sites (four inflation steps), one fifth of which is shown.
\textbf{b, } Ratio of matching dimers between both subsystems for different $\ell$ and $d$. Dimers approximately match for a large number of values for $d$, with exact matching only occurring at multiples of $L/5$ due to the pentagon tiling's $Z_5$ symmetry.
}
\label{FIG_TRANSL_INV2}
\end{figure}

Exact invariance under boundary translations, i.e., transformations \eqref{EQ_BULK_SHIFT} with $w=0$, is manifestly broken by the discretization of an $\{n,k\}$ regular hyperbolic tiling. The tiling can still be centered around either a vertex or a tile center in the Poincar\'e disk to produce a global $\mathbb{Z}_n$ or $\mathbb{Z}_k$ cyclic symmetry, but towards the boundary, these bulk rotations corresponds to an asymptotically infinite translation. 
However, we can use the same geometrical construction discussed above in the context of local scale transformations to show that boundary states still exhibit at least an approximate translation invariance under certain boundary translations.
We start from the observation in Fig.\ \ref{FIG_LOCAL_SCALING}\textbf{b} that a shift of the tiling center of a set of tiles after two inflation steps together with a deflation step, i.e., a global scale transformation, is equivalent to instead applying one or multiple deflation steps to only part of the boundary. Due to the properties of hyperbolic geometry, the non-deflated, ``joint'' part of the boundary (in the lower right of the figure) contains almost half of the edges of the deflated (red-shaded) tiles. As the boundary features (sequence of vertex types $a$ and $b$) obey a $\mathbb{Z}_5$ symmetry, i.e., are repeated five times, this implies that the vertices of the joint boundary region follow at least one fifth of the quasiperiodic sequence characterizing the boundary geometry after one inflation step, i.e., $abaab$ or any cyclic permutation thereof.
This argument holds for any bulk shift, and so many different parts of the tiling boundary after two inflation steps (blue-shaded tiles) can be identified as a joint boundary region of a shifted and deflated tiling. It follows that same sequence -- $abaab$ -- should appear in many different parts of the boundary. Of course, this is wholly expected from the quasiperiodicity of such sequences, as any inflated $a$ letter on any inflation level produces it on the next. 
Note that this does not imply an exact invariance of subsystems of a boundary state under translations preserving this boundary sequence, as the finite size of the tiling implies that long-range correlations (contributions from earlier inflation layers in the MQA) in these subsystems can still differ.
To show that short-range correlations are still closely reproduced, we consider a larger tiling after three vertex inflation steps in Fig.\ \ref{FIG_TRANSL_INV}\textbf{a}, with three shifted-and-deflated sets of tiles contained within it. This leads to three joint boundary regions labeled $A,B,C$ (not related to the previous ones) between the large and smaller sets of tiles. 
Again, we expect that these joint boundary regions are characterized by the same sequence of vertex types. Indeed, we find that $A$ and $B$ follow the sequence $ababaababa$, which appears around any inflated $a$ letter regardless of the neighboring letters as
\begin{align}
aaa &\to aba\underline{ababaababa}ab \ , \\
baa &\to \underline{ababaababa}ab \ , \\
aab &\to aba\underline{ababaabab(a)} \ , \\
bab &\to \underline{ababaabab(a)} \ .
\end{align}
This sequence appears once for every $a$ letter on the previous inflation layer. The boundary region $C$, which contains more sites than $A$ or $B$, even contains it twice, following the sequence
\begin{equation}
\underline{ababa\overline{ababa}}\overline{ababa} \ .
\end{equation}
We thus find that bulk transformations reproduce just the quasiperiodic symmetries following from the letter assignment of vertex types. Indeed, the number of such boundary translations at a given length scale is equal to the number of tiles on that inflation layer, each of which defines an allowed bulk shift (followed by a suitable rescaling that matches up boundary regions as in Fig.\ \ref{FIG_TRANSL_INV}\textbf{a}). For the $\{5,4\}$ tiling the number of tiles on a given inflation layer is equal to the number of $b$ letters on it, which implies that at a large number of inflation steps, a constant fraction 
\begin{equation}
\frac{N_b}{N_a + N_b} \to 1 - \frac{1}{\sqrt{3}} 
\end{equation}
of all boundary lattice shifts corresponds to an approximate invariance transformation.
In the MQA setup for four inflation layers, shown in Fig.\ \ref{FIG_TRANSL_INV}\textbf{b}, we see how the different boundary regions $A,B,C$ correspond to whole blocks of quasiperiodic layers. The blocks corresponding to $A$ and $B$ match on three inflation layers, while $C$ only matches the others on two layers.
As a result, subsystems of these boundary states closely match, as we demonstrate with the Majorana dimer setup in Fig.\ \ref{FIG_TRANSL_INV}\textbf{c}, where we consider the dimer subsystems corresponding to boundary regions $A,B,C$.
This self-similarity of dimer subsystems under boundary translations can also be confirmed numerically: In Fig.\ \ref{FIG_TRANSL_INV2}, we take an arbitrary dimer subsystem of size $\ell$ and consider ``dimer fidelity'' with respect to subsystems under a boundary translation $d$, that is, the percentage of matching dimers ($j \mapsto k$ mapped to $j+d \mapsto k+d$) between both states. As expected, approximately matching subsystems are found for a large range of $d$. Approximate translation invariance appears independent of $\ell$ as long as $\ell \gg 1$, showing that the quasiperiodicity is a scale-independent effect.

\subsection{Relationship to continuum CFT}

\begin{figure}[tb]
\centering
\includegraphics[height=0.26\textwidth]{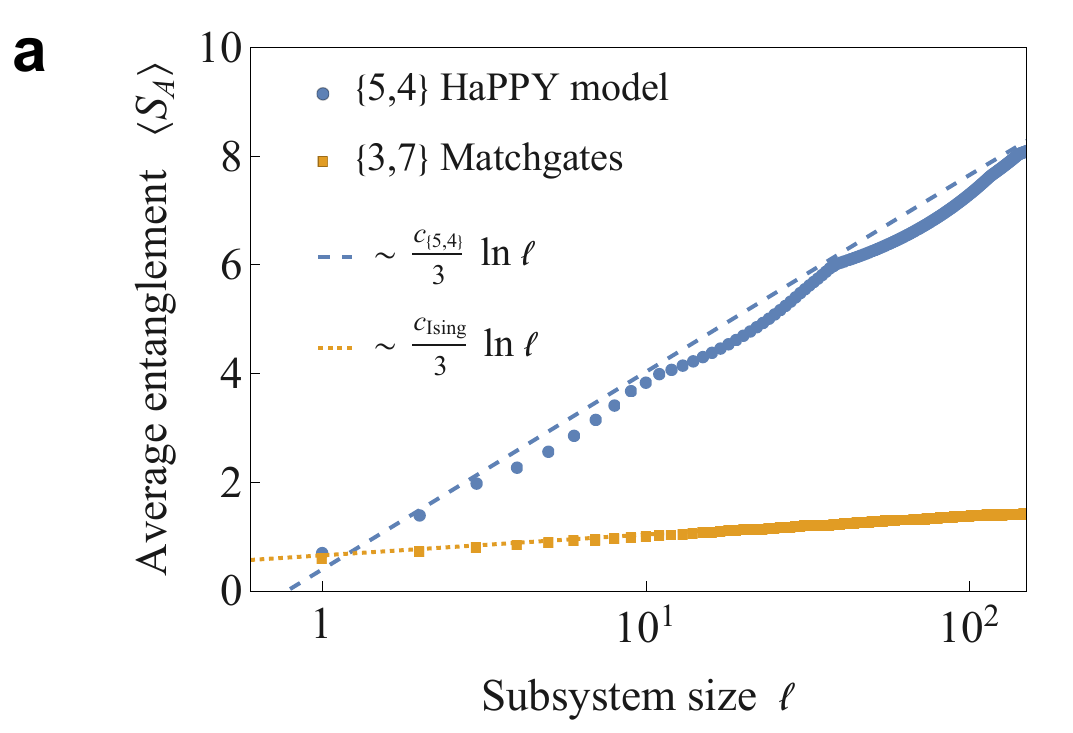}
\includegraphics[height=0.26\textwidth]{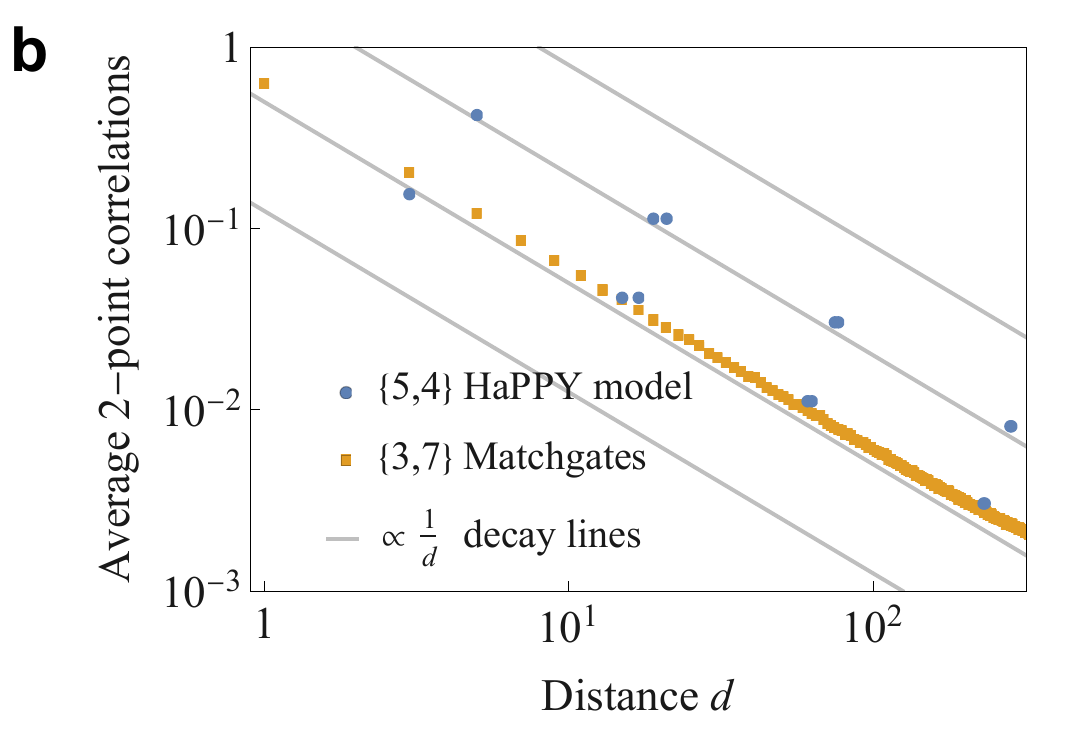}
\caption{\textbf{Physical qCFT properties. a,} Average entanglement entropy scaling for the HaPPY model (with basis state input) and matchgate tensor networks on regular geometries. Effective central charges are $c_{\{5,4\}} \approx 4.74$ and $c_\text{Ising} = \frac{1}{2}$.
\textbf{b,} Average correlation decay between Majorana modes in both models and continuum $1/d$ decay (scale arbitrary). The $\{5,4\}$ dimer result follows from the relative frequency of dimer pairs over boundary distance $d$.
}
\label{FIG_EE_AND_CORR}
\end{figure}

The symmetries of quasiperiodic CFTs have an inherent lattice structure, as they form a discrete subgroup of those of continuum CFTs. However, that does not mean that tensor networks on regular hyperbolic tiling, via whose geometry we defined qCFTs, are unable to produce states that resemble the more common regular discretizations of (ground) states of a continuum CFT. On the contrary, the enforcement of a subset of CFT symmetries is a stronger condition than other tensor network models used to describe critical states, such as the MERA, possess from geometry alone.
While some choices for tensor content and bond dimension thus result in qCFT models that clearly break CFT properties such as full translation invariance, as seen in the Majorana dimer models discussed above, others should still allow for a close approximation of continuum CFTs.
Even the dimer models can be identified with a class of CFT-like critical systems: As we saw, the $\{5,4\}$ dimer model implements an RG step that can be formulated in terms of Majorana dimer replacement rules (see details in Ref.\ \cite{Jahn:2019mbb}). However, this is merely a fermionic generalization of the RG step of \emph{strongly disordered} spin chains such as the \emph{Fibonacci XXZ chain}, where such an RG step appears in the form of a replacement rule for singlets \cite{JuhaszZimboras2007,IgloiSDRGreview}. 
This results in the same $\propto 1/d$ correlation decay with distance $d$ and $\propto \log\ell$ entanglement entropy scaling with subsystem size $\ell$ expected from a continuum CFT, as shown in Fig.\ \ref{FIG_EE_AND_CORR}, with a piecewise linear growth of entanglement characteristic for strong disorder.
Note that the average decay of correlation functions, given by a histogram of the dimers/singlets over the boundary distance $d$ over which they are paired up, appears ``split'' into two series in the dimer case; this is the result of two types of dimer pairs that appear at each length scale (compare Fig.~\ref{FIG_TRANSL_INV}\textbf{c}). 
A second, more realistic qCFT model to consider is that of generic tensor networks on a regular $\{3,7\}$ geometry using \emph{matchgate tensors} as studied in Ref.\ \cite{Jahn:2019nmz}. These tensors can encode any fermionic Gaussian state and can be contracted with a computational effort that is only polynomial, rather than exponential, in the number of contracted bonds.
For the simplest instance of this model at bond dimension $\chi=2$, correlation functions with only soft deviations from translation invariance and site-averaged properties (scaling dimensions and OPE coefficients) matching the continuum Ising CFT can already be recovered \cite{Jahn:2019nmz}. Entanglement scaling and two-point correlation decay is shown in Fig.\ \ref{FIG_EE_AND_CORR} along the results for the dimer model. In particular, the entanglement entropy smoothly follows the Calabrese-Cardy prediction \cite{CalabreseReview}, rather than the piece-wise linear growth of the disordered Majorana dimer states.
This model illustrates that qCFT properties can closely approximate continuum CFTs even at small bond dimension. To recover full translation invariance, tensor networks at larger bond dimension become necessary. Such a construction in terms of matchgate tensors will be explored in future work.
As a third example, consider the setup of \emph{hyper-invariant} tensor networks \cite{PhysRevLett.119.141602} also based on vertex inflation on regular hyperbolic tilings, using tensors with certain constraints to reproduce properties of MERA-type tensor networks.  In particular, these symmetries preserve the full tiling symmetries and thus also classify as an instance of a qCFT. Indeed, it has recently been shown that this ansatz also leads to boundary states that are consistent with certain conformal properties \cite{Steinberg:2020bef}, confirming that the class of qCFTs includes at least close approximations of continuum CFTs.

\section{Discussion}

In this work, we found that tensor networks on regular $\{n,k\}$ tilings with equivalent tensors on all sites lead to boundary states that respect a discrete subset of conformal symmetries, specifically invariance under local and global scale transformations, as well as an approximate translation invariance.
Due to the associated quasiperiodic boundary symmetries, we call a theory invariant under these transformations a \emph{quasiperiodic conformal field theory} (qCFT) in analogy to the symmetry definition of a continuum CFT.
As in the continuum, states that are invariant under all discrete qCFT transformations are associated with ground states, while excited states need only respect the asymptotic symmetries at small scales. In the bulk perspective, this corresponds to deforming either the regular geometry or changing the tensor content around the center of the tiling while leaving the near-boundary structure unchanged.
Our approach expands upon the idea of \emph{conformal quasicrystals} \cite{Boyle:2018uiv} which posits that tensor networks on regular hyperbolic tilings describe boundary states on a quasiperiodic lattice, but differs from it in several respects: Here we argued that the combined symmetries of all tiling layers characterize the final boundary states, rather than the boundary symmetries alone, as captured by our multi-scale quasicrystal ansatz (MQA). 
Physical models with local couplings that are quasiperiodically modulated between two or more values, such as the Fibonacci XXZ model \cite{IgloiSDRGreview}, more properly describe spin chains on quasicrystals but have symmetries that are simpler than that of the qCFT framework described here.
Furthermore, we related discretized bulk symmetries to specific boundary conformal transformations, whereas Ref.\ \cite{Boyle:2018uiv} considered deformations of the initial tiling before inflation as defining such transformations. However, such deformations only change the long-range (IR) symmetries of the boundary state in the manner of a relevant CFT operator and thus cannot describe the full ground state symmetries. In addition, we found that the choice of inflation rule changes the effective central charge of the boundary state, which leads to the restricted class of ``allowed'' discretized symmetry transformation described here.
We expect qCFTs to appear on the boundary of all tensor network models on a regular bulk geometry \cite{Pastawski2015,PhysRevA.98.052301,Osborne:2017woa,PhysRevLett.119.141602,Hayden2016}, potentially including p-adic AdS/CFT models \cite{Gubser:2016guj,Heydeman:2016ldy}.
From their construction on regular tilings, these models necessarily inherit the discretely broken conformal symmetries we discussed, thus falling into the rubric of our proposed AdS/qCFT correspondence.
The HaPPY pentagon code, studied here in its Majorana dimer form, shows that qCFTs are closely related to strongly disordered critical models, which have been extensively studied in the condensed matter literature \cite{PhysRevB.51.6411,PhysRevLett.93.260602,IgloiSDRGreview,PhysRevX.5.031032,Tsai_2020,PhysRevX.10.011025}.
As the HaPPY code is a model of quantum error correction, we find that AdS/qCFT includes exact holographic codes, rather than the approximate codes found in AdS/MERA models \cite{Kim2016}.
The similarity to strongly disordered models also suggests that boundary dynamics can be described by an effective \emph{local} (though not nearest-neighbor) Hamiltonian, which would allow for dynamical AdS/qCFT models.
Further work should also explore the role of qCFT excitations, where the regular symmetries of the tensor network are (locally) broken.
On a mathematical side, a rigorous extension of the discrete breaking of $PSL(2,\mathbb{R})$ symmetries through Fuchsian groups to the breaking of full conformal symmetries by space-time lattices would help in the construction of new tools to study time evolution in CFTs.
Similar notions of discrete conformal transformations should also appear when considering the geometry of higher-dimensional regular hyperbolic tilings.

\section{Acknowledgements}
We thank Marek Gluza, Xiaoliang Qi, Sukhbinder Singh, 
Tadashi Takayanagi, and Charlotte Verhoeven for helpful comments and discussions. This work has been supported by the Simons Collaboration on It from Qubit, the Templeton Foundation, the DFG (CRC 183, EI 519/15-1), and the FQXi. This  work  has  also  received  funding  from  the  European  Union's Horizon 2020 research and innovation programme under grant agreement  No.~817482 (PASQuanS).
This research has been supported in part by the Perimeter Institute for Theoretical Physics. Research at Perimeter Institute is supported by the Government of Canada through the Department of Innovation, Science, and Economic Development, and by the Province of Ontario through the Ministry of Research and Innovation. We also acknowledge support from the National Research, Development and Innovation Office (NKFIH) through the Quantum Information National Laboratory of Hungary and  Grants No. K124176, FK135220, K124351.

\end{document}